\documentclass[10pt,letterpaper]{article}
\usepackage{opex3}
\usepackage{cite}

\begin{document}

\title{Precise measurement of laser power using an optomechanical system}

\author{Kazuhiro Agatsuma,$^{1, 2, *}$ Daniel Friedrich,$^2$ Stefan Ballmer,$^3$ \\Giulia DeSalvo,$^4$ Shihori Sakata,$^2$ Erina Nishida,$^5$ \\and Seiji Kawamura$^2$}

\address{
$^1$ National Astronomical Observatory of Japan, Mitaka, Tokyo 181-8588, Japan\\
$^2$ Institute for Cosmic Ray Research, University of Tokyo, Kashiwa, Chiba 277-8582, Japan\\
$^3$ Department of Physics, Syracuse University, NY 13244, USA\\
$^4$ University of California, Berkeley, CA 94720, USA\\
$^5$ Ochanomizu University Graduate School of Humanities and Sciences, Bunkyo, Tokyo 112-8610, Japan\\}
\email{*agatsuma@nikhef.nl} 



\begin{abstract}
This paper shows a novel method to precisely measure the laser power using an optomechanical system. 
By measuring a mirror displacement caused by the reflection of an amplitude modulated laser beam, 
the number of photons in the incident continuous-wave laser can be precisely measured. 
We have demonstrated this principle by means of a prototype experiment uses a suspended 25\,mg mirror as an mechanical oscillator coupled with the radiation pressure 
and a Michelson interferometer as the displacement sensor. 
A measurement of the laser power with an uncertainty of less than one percent (1\,$\sigma $) is achievable. 
\end{abstract}



\section{Introduction}
Optical power meters used to measure the laser power are among the essential tools for optics experiments nowadays. 
The absolute value of the laser power was originally measured using phototubes that are based on the photoelectric effect \cite{Maiman1961}. 
Subsequently, 
calorimeters were used to determine the laser power (even for the masers) by measuring a temperature increase that corresponds to absorbed laser energy \cite{Li1962}. 
The semiconductors and thermopile sensors, 
which are based on conversions of radiation into the electrical current and of thermal gradient into the electrical current, respectively, 
are applied for optical power measurements, but they require a calibration referring to a primary standard. 
Calorimeters continue to be developed and have defined the primary standards of the laser power and energy with a traceability to SI units for nearly half a century. 
The current primary standard is based on a family of isoperibol calorimeters \cite{West1970}. 
These standards have a 0.25\,\% uncertainty with 1\,$\sigma $ (it is a probability of 68\,\% that the true value is included within its uncertainty), 
on broadband wavelengths (193\,nm - 10.6\,$\mu $m) \cite{NIST}. 
The world best measurements, as long as we know, 
were performed using cryogenic radiometers under some limited wavelengths and powers with an associated uncertainty of 0.01\,\% (1\,$\sigma $) \cite{Livigni1998, POWR}. 
Although typical commercial radiometers (power meters) refer to these primary standards, they still have an uncertainty of few percent. 

Recently, the application range of power meters is spreading, so that a power measurement has become relevant to the quality of implemented experiments.
Free electron lasers, for example, have a broad range of wavelength (around 0.1\,nm - 1000\,$\mu $m) \cite{FEL, FEL2011}. 
Therefore, it is interesting to develop primary standards for such wavelengths (see e.g. \cite{Kato2012}). 
In addition, 
photodiodes (PDs) with high quantum efficiency (QE) (close to 99\,\%) are required for optical experiments that can improve their sensitivity by means of squeezed light, 
such as gravitational wave (GW) detectors \cite{Kimble2002, Vahlbruch2008, GEO2010}.
The QE of a PD is determined by measuring the incident laser power. 
Hence, the uncertainty of QE measurement is limited by the power measurement, i.e. its precise measurement (below one percent) is essential. 
Also, 
photon pressure calibrator, which is used for a displacement calibration of GW detectors, 
requires a precise value of the laser power as a reference of the calibration \cite{Clubley2001, Mossavi2006}. 

We propose a new technique to measure the laser power using an optomechanical-coupled oscillator, 
which allows to determine the laser power from a displacement measurement. 
In addition, we have demonstrated this technique by constructing a prototype to investigate its feasibility. 
The derived uncertainty is below one percent (1\,$\sigma $), which is better than that of current commercial power meters. 
Our apparatus can be applied for arbitrary laser wavelengths by adapting material and coating of the oscillator. 
This technique has the potential to be an alternative method to realize a primary standard. 
First, theory and concept are shown. 
Subsequently, experimental results are discussed, 
which is followed by a detailed analysis of uncertainties that contribute to the overall measurement uncertainty.  

\section{Theory} 
\subsection{Concept} 
\begin{figure}[htbp]
 \centering
      \includegraphics[width=8cm ]{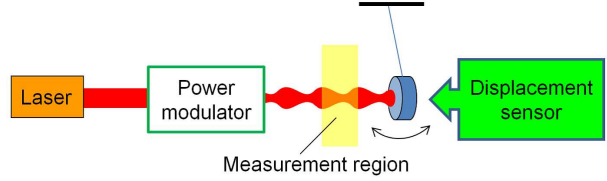}
         \caption{Sketch of the optomechanical system investigated in this work. 
By measuring the displacement of a suspended mirror the incident modulated laser power can be determined.
}
         \label{fig:Concept}
\end{figure}

For an optomechanical system as sketched in Fig.\,\ref{fig:Concept} with an input laser and a suspended mirror coupled through the radiation pressure, 
the laser power can be related with displacement of the mirror shaken by input photons. 
The displacement of a mirror pushed by radiation pressure in the frequency domain can be expressed as \cite{Clubley2001}
\begin{equation}
 d \tilde{X} = \frac{ 2 P_\mathrm{m} }{ c \, m \, \omega ^2 }.  \label{eq:Opmech}
\end{equation}
Here, $ d \tilde{X} $ is the Fourier transformation of the displacement of the mirror, 
$ m $ the mass of the mirror, 
$ c $ the speed of light, 
$ P_\mathrm{m} $ the intensity-modulated laser power, 
and 
$ \omega $ the angular frequency of the modulation. 
In Eq.\,(\ref{eq:Opmech}), the mirror response is regarded as the free-oscillation mass (`free mass'). 
The modulated laser power incident on the mirror can be derived by measuring the displacement of the moving mirror at the modulation frequency. 

The procedure to measure the DC laser power requires a characterization of the modulation strength at the measurement frequency. 
For this purpose, a reference receiver like a PD is used to determine the ratio of the DC power and the modulated power. 
If the PD is put in the measurement region (see Fig.\,\ref{fig:Concept}) after obtaining a value of the modulated laser power, 
the output voltage of the PD is related to the modulated power. 
The ratio of them provides $ C_\mathrm{m} \equiv  P_\mathrm{m}/V_\mathrm{m} $; 
$ C_\mathrm{m} $ is the conversion factor between the voltage and laser power,   
$ V_\mathrm{m} $ is the spectrum value of the output voltage of the PD at the modulation frequency. 
If the PD response is independent of the frequency, the conversion factor is used for the signal not only at the modulation frequency but also DC. 
A well calibrated PD can be used as a precise power meter, for even other purposes. 

For a known mechanical response of the mirror as well as a known power modulation one can actuate the mirror by a well defined amount. 
This is known as a photon pressure calibrator and is, for example, used in GW detectors \cite{Clubley2001, Mossavi2006}. 
On the other hand, for a measured displacement of the mirror one can instead derive the modulated laser power, 
which is the concept of the novel power meter investigated in this work. 
In contrast to GW detectors, we use a lightweight mirror in the order of 25\,mg in order to increase the optomechanical response in the experiment. 
As a displacement sensor we have used a Michelson Interferometer as explained in the following section. 

\subsection{Theoretical model} 
In the following, it is outlined how the incident laser power can be derived from the displacement of a suspended mirror. 
This includes a brief discussion of the optomechanical model for a quasi free mass as well as of the readout via a Michelson interferometer, 
which is the basis of the investigated concept. 
Further information can be found in section\,\ref{Sec:Uncertainty}, where the uncertainty contribution of each parameter is discussed in detail.  

Generally, the incident laser can be reflected at the mirror under non-normal incidence and off center. 
These two effects can be accounted for by modifying Eq.\,(\ref{eq:Opmech}) yielding off centering of the beam spot on the mirror. 
Also, imperfectness of the reflectivity of the mirror should be taken into account. 
Thus, the optomechanical response becomes 
\begin{equation}
 d\tilde{X} = \frac{ 2 P_\mathrm{m} \alpha _{\mathrm{r}} \cos{\phi } }{ c \, m \, \omega ^2 } (1 + R_\mathrm{c}),                           \label{eq:Disp_photon}
\end{equation}
where, 
$ \alpha _{\mathrm{r}} $ is the transfer efficiency of the momentum from photons to the mirror motion, 
$ \phi  $ the incident angle of the input laser to the mirror 
and 
$ R_\mathrm{c} $ the rotational effect from off centering of the beam spot on the mirror. 
Details about the latter effect can be found in section\,\ref{Sec:Rc}. 
If the mirror is suspended by wires, the response of the mirror is regarded as that of a free mass for frequencies much higher than the pendulum resonance frequency. 
However, the response of the mirror has a small deviation from the perfect free mass even in such frequency region.
A more general expression uses $ H_{\mathrm{m}} (\omega ) $ as the mechanical response below 
instead of the free mass expression $ 1/(m \omega ^2) $ in Eq.\,(\ref{eq:Disp_photon}).

We selected the Michelson interferometer (MI) as a displacement sensor because of its simplicity and a suitable sensitivity. 
When the so-called `mid-fringe' state is assumed, the readout signal is related with the displacement via 
\begin{equation}
 d\tilde{X} = ( \lambda \big{/} 2 \pi V_{\mathrm{pp}} ) dV_{\mathrm{PD}} ,           \label{eq:PD_MI}
\end{equation}
where $ dV_{\mathrm{PD}} $ is the output voltage from the PD 
and $ V_{\mathrm{pp}} $ is the peak-to-peak voltage when the arm length is changed over half a wavelength $ \lambda $. 
Combining Eq.\,(\ref{eq:Disp_photon}) and Eq.\,(\ref{eq:PD_MI}) yields
\begin{equation}
 P_\mathrm{m} = \frac{ c \lambda }{ 4 \pi V_{\mathrm{pp}} H_{\mathrm{m}} \alpha _{\mathrm{r}} (1 + R_\mathrm{c}) \cos{\phi }  } \, dV_{\mathrm{PD}} .     \label{eq:Power}
\end{equation}
Note that if the mid-fringe state is kept by position control, $ dV_{\mathrm{PD}} $ with activated control loop is different from the one without control loop. 
\begin{figure}[htbp]
 \centering\includegraphics[width=6.5cm ]{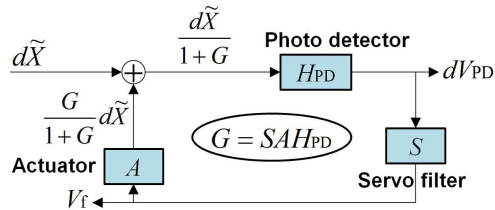}
         \caption{Block diagram for the MI control. 
The symbols $ S, A $ and $ H_{\mathrm{PD}} $ show each transfer function of the servo filter, actuator response and PD response, respectively.
}
         \label{fig:Blockdiagram}
\end{figure}
To include the effect of the control loop, 
the term $ dV_{\mathrm{PD}} $ in Eq.\,(\ref{eq:Power}) has to be replaced by $ V_{\mathrm{f}} G_{\mathrm{CL}} T_{\mathrm{AH}} $ 
because $ d\tilde{X} = V_{\mathrm{f}} G_{\mathrm{CL}} T_{\mathrm{AH}} / H_{\mathrm{PD}} $ (see Fig.\,\ref{fig:Blockdiagram}).
Here, 
$ V_{\mathrm{f}} $ is the feedback signal of the MI, 
$ G_{\mathrm{CL}} = (1+G)/G $ the closed-loop gain, 
$ G $ the loop gain 
and
$ T_{\mathrm{AH}} $ the transfer function from actuator to PD. 
Finally, we obtain  
\begin{equation}
 P_{\mathrm{m}} = \frac{ c \lambda \, V_{\mathrm{f}} G_{\mathrm{CL}}  T_{\mathrm{AH}} }
                                { 4 \pi V_{\mathrm{pp}} H_{\mathrm{m}} \alpha _{\mathrm{r}} (1 + R_\mathrm{c}) \cos{\phi }  }  .   \label{eq:Power_2}
\end{equation}
The absolute values of each transfer function are used in Eq.\,(\ref{eq:Power_2}). 
We focus on the uncertainty of the power measurement in this paper. 
The uncertainty in $ \lambda $ is included in the uncertainty evaluation of $ V_{\mathrm{f}} $. 
The displacement readout part ($ V_{\mathrm{f}} $, $ G_{\mathrm{CL}} $, $ T_{\mathrm{AH}} $ and $ V_{\mathrm{pp}} $) is evaluated 
by means of a prototype experiment to estimate actual effects, which are difficult to evaluate by just calculation, 
from electrical noise, laser intensity noise, thermal drift and so on. 
The optomechanical response part ($ H_{\mathrm{m}} $, $ \alpha _{\mathrm{r}} $, $ R_\mathrm{c} $ and $ \phi $) is analyzed using both modeling and measurements. 
A detailed discussion is given in section\,\ref{Sec:Uncertainty}. 

\section{Experimental setup}
\begin{figure}[htbp]
 \centering\includegraphics[width=\linewidth ]{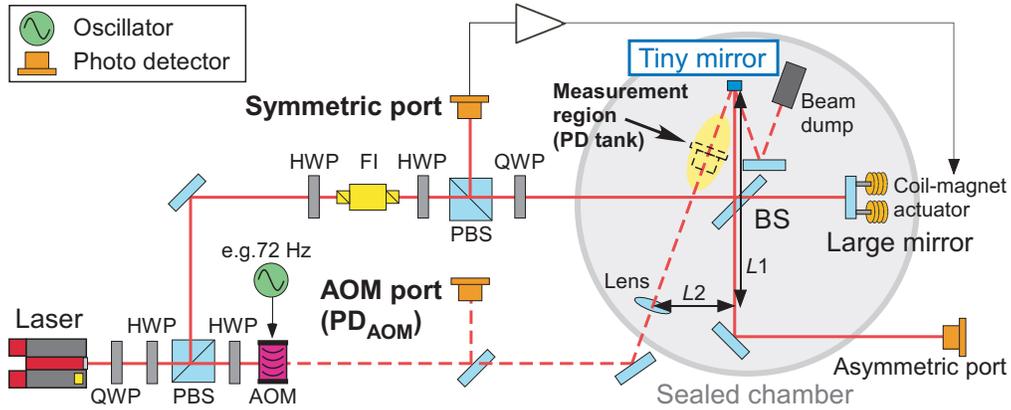}
         \caption{Schematic view of the experimental setup.
Acronym explanation: 
HWP, half wave plate; 
QWP, quarter wave plate; 
BS, beam splitter; 
PBS, polarized beam splitter; 
FI, faraday isolator; 
and AOM, acousto-optic modulator. 
$ L1 $ and $ L2 $ show length measurements for the incident angle.
There are three photo detectors; Symmetric port, Asymmetric port and AOM port on an optical table.
The solid red line indicates a beam trace for the MI and the dotted red line is that for an intensity modulation by AOM. 
}
         \label{fig:Setup}
\end{figure}

When a light weight mirror is used, the displacement due to the radiation pressure of the incident light is increased (see Eq.\,(\ref{eq:Disp_photon})). 
We use a tiny flat mirror of about 25\,mg as the optomechanical-coupled oscillator. 
The displacement sensor is a MI that consists of the tiny mirror and a one-inch flat mirror (`large mirror') as end mirrors in the two arms, respectively. 
A schematic view of the experimental setup is shown in Fig.\,\ref{fig:Setup}. 
The laser (Innolight Inc. Mephisto) has a nominal power of 500\,mW and a wavelength of 1064\,nm. 
The laser beam is divided into two paths that are used for the Michelson interferometer ($\sim$50\,mW) and to actuate the tiny mirror ($\sim$450\,mW). 
The latter one is modulated in power by means of an acousto-optic modulator (AOM), which provided a power modulation index of about 17\,\%. 
In the MI light path, the laser power that remains at the front of the tiny mirror is about 10\,mW. 
The MI is housed in a sealed chamber, which is put on vibration isolation stacks, to be soundproof in air. 
A preliminary experiment can be found in reference \cite{Agatsuma2012}. 

\begin{figure}[htbp]
 \centering\includegraphics[width=\linewidth ]{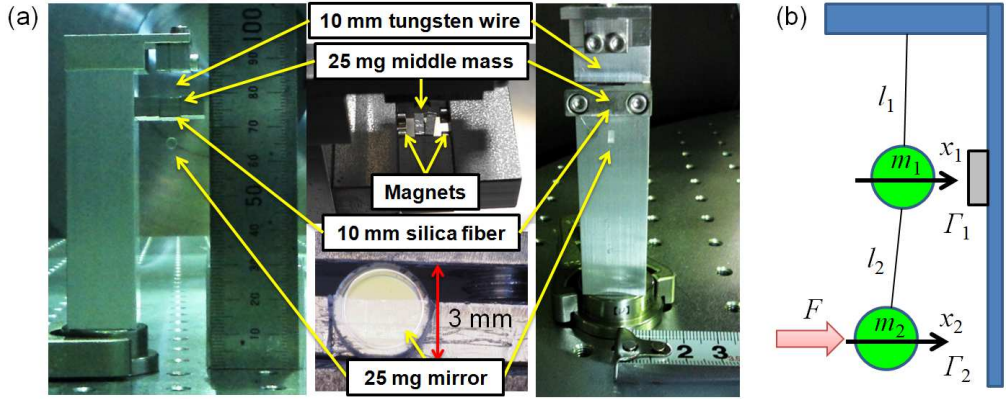}
         \caption{The tiny mirror suspension system. (a) photographs of the 25\,mg mirror and the double pendulum. (b) model of the double pendulum.}
         \label{fig:DoublePend}
\end{figure}
Figure\,\ref{fig:DoublePend}(a) shows photographs of the small suspension system. 
The tiny mirror has a cylindrical shape with a diameter of 3\,mm, a thickness of 1.5\,mm and is suspended as a double pendulum developed in \cite{Sakata2010}. 
The lower suspension is a silica fiber with a diameter of 10\,$ \mu $m, glued on the mirror using UV cured resin. 
The middle mass of 25\,mg, made of aluminum, is suspended by a 10\,$ \mu $m tungsten wire 
and is eddy current damped by the surrounding small magnets. 
The second mirror of the MI is a conventional one inch mirror with a flat surface and a weight of 48\,g (including optical bracket). 
This suspension system is also a double pendulum with eddy current damping. 
The large mirror and its optical bracket are suspended by 50\,$ \mu $m tungsten wires. 
The intermediate mass of 57\,g is suspended by 70\,$ \mu $m tungsten wires. 

The large mirror is actuated to keep the MI on its operation point by controlling the differential length of the two interferometer arms. 
This feedback control is called `lock'. 
For this purpose, four sets of coil-magnet actuators are used for position control of the large mirror.
The error signal for the mid-fringe lock is derived by applying an electrical offset to the PD output at the symmetric port. 
The offset is adjusted to provide a zero crossing at the center of the fringe. 
A linear response for the displacement of the mirror is guaranteed as long as the mid-fringe state is maintained. 
A contrast of about 95\,\% is kept in this experiment. 

\section{Experimental results}
\subsection{Sensitivity and radiation pressure response}
\begin{figure}[htbp]
 \centering\includegraphics[width=\linewidth ]{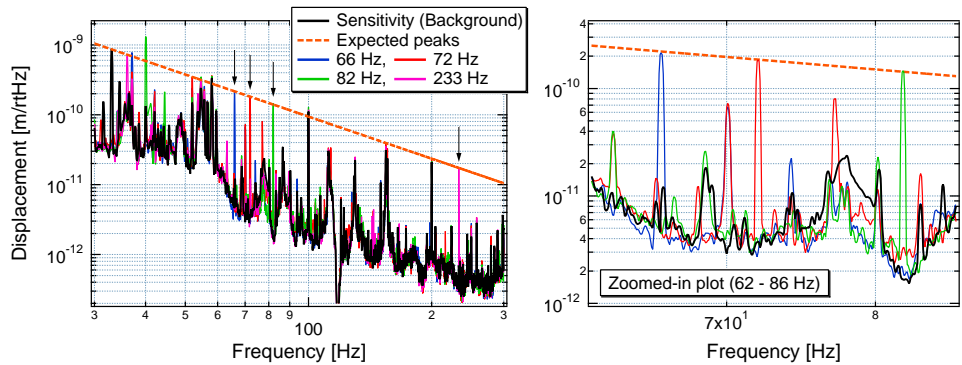}
 \caption{The sensitivity of the MI and the displacement shaken by radiation pressure at 66, 72, 82 and 233\,Hz.
The orange dashed line shows $f ^{-2}$ response of the free mass from Eq.\,(\ref{eq:Disp_photon}).
} 
 \label{fig:Sens}
\end{figure}
In Fig.\,\ref{fig:Sens}, apparent peak signals due to the power modulation are measured together with  
the displacement of the MI without power modulation, which is called the sensitivity (background). 
Four frequencies (66, 72, 82 and 233\,Hz) have been chosen for the amplitude modulation because a sufficient signal to noise ratio is provided. 
Furthermore, they show a $f ^{-2}$ response (orange dashed line) as it is predicted for a free mass 
(a quasi free mass, in a rigorous expression as discussed in section\,\ref{Sec:Hm}). 
In our measurement, the spectra and standard deviations are obtained from the time series data of 1\,kHz sampling with an anti-aliasing filter of 500\,Hz.
The flat-top window is used when the Fourier transformation is applied to the data to evaluate precise values of the peak structure.

\subsection{Laser power measurement}
The modulated laser was partially extracted for the PD on the AOM port ($ \mathrm{PD_{AOM}} $) instead of on the measurement region. 
Here, $ C_\mathrm{m}^* = P_\mathrm{m} / V_{\mathrm{PDAOM}} $ becomes the corresponding conversion factor. 
Once $ C_\mathrm{m}^* $ is determined, $ \mathrm{PD_{AOM}} $ can be a real time monitor of the laser power at the front of the small mirror. 
A high power beyond a PD receivable can also be sent to the measurement region by utilizing this $ \mathrm{PD_{AOM}} $. 
It means we can see the response signal with reinforcing the optomechanical coupling. 
The conversion factor $ C_\mathrm{m}^* $ is a constant as long as the condition of the path of the modulated beam is kept. 
Hence, the optical components in the beam path are not related with any uncertainty contribution. 
If power loss is caused by mirror contamination, $ C_\mathrm{m}^* $ should be measured again. 
The results presented in the following are based on measurements with a modulation frequency of 72\,Hz, as it showed the lowest uncertainty in our setup.

The modulated power $ P_\mathrm{m} $ is calculated using Eq.\,(\ref{eq:Power_2}) and the peak value at 72\,Hz in Fig.\,\ref{fig:Sens}. 
The result was $ P_\mathrm{m} = 39$\,mW, which corresponds to $ 2.1 \times 10^{17} $ photons. 
The conversion factor of 0.371\,W/V was obtained by dividing the value of modulated power by the output voltage of $ \mathrm{PD_{AOM}} $ at 72\,Hz. 
The resultant DC voltage of the $ \mathrm{PD_{AOM}} $ was 0.802\,V (environmental offset was subtracted) when the modulation of AOM was turned off, 
which corresponds to the laser power of 298\,mW at a close position from the tiny mirror. 
Note that the above estimation did not include the rotational effect as shown in Eq.\,(\ref{eq:Power_2}) by $ R_\mathrm{c} $. 

\subsection{Quantum efficiency measurement}
\label{Sec:QE}
As an application of this power measurement, the QE can be estimated as explained in the following. 
The QE of a PD is defined as the conversion efficiency from the light power to the current; $ QE = N_\mathrm{e} / N_\mathrm{P} $. 
Here,  $ N_\mathrm{e} $ is the number of electrons in the output current of the PD and $ N_\mathrm{P} $ the number of photons in the input light. 
$ N_\mathrm{P} $ is related to a continuous-wave laser with power $ P $ by $ N_\mathrm{P} = P / ( h \nu ) $ 
using Planck's constant $ h $ and the frequency of the photon $ \nu $. 
This QE corresponds to the inverse of the conversion factor ($ C_\mathrm{m}^{-1} $) by multiplying the inverse of an inner resistance $ R $ of the PD. 
This implies that a precise QE of the PD can also be obtained using our power meter. 
Note that the relation of $ QE = (C_\mathrm{m} R)^{-1} $ can be used when the PD is calibrated in the measurement region. 
To derive the QE of the $ \mathrm{PD_{AOM}} $, we need the laser power ratio between the AOM port and the measurement region. 

After calibrating the $ \mathrm{PD_{AOM}} $ as a power meter, another PD (PD tank) to measure the QE was put in the measurement region. 
Usually, PDs cannot receive the incident power over 10\,mW. 
A neutral density filter is inserted close to the AOM output in order to reduce the laser power. 
Since the $ \mathrm{PD_{AOM}} $ can monitor such a change of the power including a drift, 
as long as the power linearity of the PD is guaranteed, 
the incident laser power at the measurement region can be obtained with the same uncertainty as the above calibration. 
The measured value of the QE of the PD tank is 0.301, corresponding to the photo sensitivity of 0.258\,A/W.
In order to also measure the QE of $ \mathrm{PD_{AOM}} $, the positions of both PDs were exchanged. 
The beam spot sizes on the PD surfaces are adjusted to be almost same size (0.2\,mm) by measuring beam profiles and adjusting the PD positions, 
in case the response of the PDs change by the beam spot size. 
Comparing with the $ \mathrm{PD_{AOM}} $, the PD tank has an efficiency ratio of 0.939 and a laser power ratio of 102.
The $ \mathrm{PD_{AOM}} $ has a QE of 0.320, corresponding to 0.275\,A/W. 
Both PDs are Si PIN photodiode (S3759 produced by HAMAMATSU). 

According to the specification sheet, the expected photo sensitivity is the range from 0.3 to 0.38\,A/W, which comes from individual differences.
The results are lower than the specification sheet by 20\,\%, at least. 
The reason seems the unmeasured rotational effect, corresponding to an off centering of about 0.4\,mm, 
and low temperature in the laboratory (about 10 degree lower than that of the specification sheet).
Also, note that the beam position on the PD surface should be considered to achieve more careful measurements. 
In order to obtain an accurate value of the QE, above items should be treated quantitatively. 
Using this method, the QE can be measured precisely in principle. 
Now, our focus is to evaluate the inevitable uncertainty of the power measurement, which is analyzed in the following section. 

\section{Uncertainty evaluation}
\label{Sec:Uncertainty}
\subsection{Definition and method}
We use the standard uncertainty to evaluate each measurement. 
According to a document of ISO \cite{GUM2008}, 
the standard uncertainty is defined as uncertainty of the result of a measurement expressed as a standard deviation.
There are two ways of defining the standard uncertainty, the so-called Type A and Type B evaluation.
Type A evaluation is used for the statistical uncertainty, which has a Gaussian distribution giving the standard deviation.
Type B evaluation is used for the results that are not obtained from repeated observations, e.g. an upper and lower limit.
Since the uniform probability density function is used under Type B evaluation, 
the uncertainty is divided by $ \sqrt{3} $ to correspond to 1\,$\sigma $ of the standard deviation. 

The propagation law of uncertainty (combined standard uncertainty $ \sigma _Y $) is expressed as 
\begin{equation}
 \sigma _Y = \sqrt{ 
            \sum_{i=1}^{s} \left( \frac{\partial Y}{\partial \xi_i} \sigma_i \right) ^2 + 
            2 \sum_{i=1}^{s-1} \sum_{j=i+1}^{s}  \frac{\partial Y}{\partial \xi_i} \frac{\partial Y}{\partial \xi_j} \sigma_i \sigma_j r( \xi_i, \xi_j ) ,
}           
\label{eq:Sigma}
\end{equation}
where 
$ Y $ is a certain complex function that consists of $ \xi_i $ ($ i = 1, 2, .., s $), 
$ \sigma_i $ the standard uncertainty of each component 
and $ r( \xi_i, \xi_j ) $ the estimated correlation coefficient associated with $ \xi_i $ and $ \xi_j $. 
Using both Eq.\,(\ref{eq:Power}) and Eq.\,(\ref{eq:Sigma}), the total uncertainty of the laser power is written as 
\begin{equation}
 \frac{\sigma _P}{P} = \sqrt{ 
\left( \frac{\sigma _{dV}}{dV_{\mathrm{PD}}} \right) ^2 + 
\left( \frac{\sigma _{V} }{ V_{\mathrm{pp}} } \right) ^2 + 
\left( \frac{\sigma _H}{H_{\mathrm{m}}} \right) ^2 + 
\left( \frac{\sigma _\alpha}{\alpha_{\mathrm{r}} } \right) ^2 + 
\left( \sigma _\phi \tan \phi \right) ^2 + 
\left( \frac{\sigma _R}{ 1 + R_\mathrm{c} } \right) ^2 .
}                                                                           \label{eq:Sigma_P}
\end{equation}
It is assumed that ingredients of uncertainty have no correlation to each other so that the second term of Eq.\,(\ref{eq:Sigma}) vanishes. 
In order to keep the uncertainty of the laser power within one percent, 
the relative uncertainty of below 0.3\,\% is required 
because the number of effective components is about ten when taking into account the different contributions included in the first term of Eq.\,(\ref{eq:Sigma_P}).
Here, the relative uncertainty is a ratio between each component and its standard deviation ($\sigma _i / \xi_i $). 
Hence, the relative uncertainty of 0.3\,\% is a naive requirement (criterion) and of course a lower value is desirable. 

\subsection{Deviation from mid-fringe lock $ dV_{\rm{PD}} $ }
\begin{figure}[htbp]
 \begin{center}
      \includegraphics[width=\linewidth ]{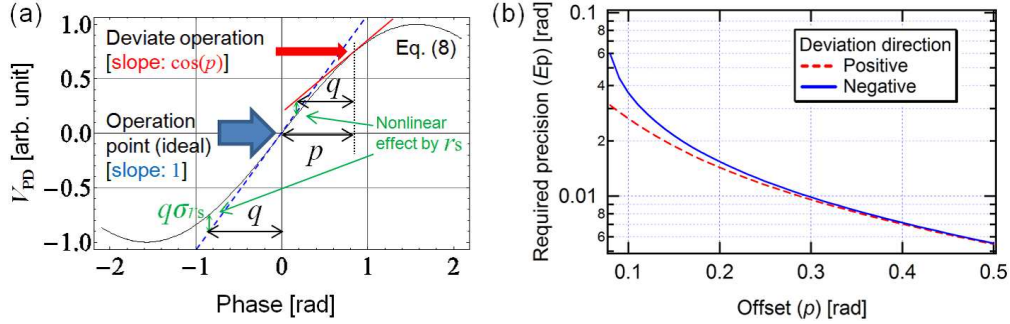}
         \caption{Deviation from the mid-fringe lock. 
(a) effect of $ k $ and $ r_\mathrm{s} $. 
Here, 
$ p $ is an offset phase of an operation point, 
$ q $ an average phase of a residual error signal and 
$ r_\mathrm{s} $ the difference between an ideal linear response and an actual sinusoidal response; 
(b) required precision for the measurement of p from Eq.\,(\ref{eq:Ep}). 
}
         \label{fig:TiltCorr}
  \end{center}
\end{figure}

The voltage at the error point obtained from the PD at the symmetric port of the MI depends on the arm length difference $x$ and can be written as 
\begin{equation}
  V_{\rm{PD}} = \frac{1}{2} V_\mathrm{pp} \sin{ \left[ 2 \pi \Big( \frac{x}{\lambda /2} \Big) \right] }  .    \label{eq:PD_MI_sin}
\end{equation}
The PD response $ H_{\rm{PD}} $ is produced by its derivative. 
If a perfect condition of the mid-fringe ($ x = \lambda /8 + \lambda n/2 $) ($ n $ is integer) is kept, 
Eq.\,(\ref{eq:PD_MI}) is yielded using relations of $ dV_{\rm{PD}} = H_{\rm{PD}} d\tilde{X} $ and $ H_{\rm{PD}} = 2 \pi V_\mathrm{pp} / \lambda $.
We used an electrical offset to adjust the operation point of the interference to the mid-fringe.
The perfect mid-fringe is desirable because a linear response of the MI is obtained around that point. 
However, the electrical circuit may cause a drift of the offset, 
changing the gradient of linear response due to the term of $ \cos(p) $ in the derivative of Eq.\,(\ref{eq:PD_MI_sin}) 
where $ p $ is phase of offset (see Fig.\,\ref{fig:TiltCorr}(a)). 
In addition, a large residual error signal cause a nonlinear effect because of actual sinusoidal response in Eq.\,(\ref{eq:PD_MI_sin}). 
Namely, the PD response is rewritten as 
\begin{equation}
  H_{\rm{PD}} = \frac{2\pi V_\mathrm{pp}}{\lambda } k r_\mathrm{s} ,  \label{eq:H_PD}
\end{equation}
where $ k = \cos{(p)} $ and $ r_\mathrm{s} $ is difference between the sinusoid and linear. 
The uncertainty term from the PD response is expanded 
with the relation of $ dV_{\mathrm{PD}} = V_{\mathrm{f}} G_{\mathrm{CL}} T_{\mathrm{AH}} $ including deviation terms $ k $ and $ r_\mathrm{s} $ as 
\begin{equation}
\left( \frac{\sigma _{dV} }{ dV_{\mathrm{PD}} } \right) ^2 
=
  \left( \frac{\sigma _{V_\mathrm{f} } }{V_\mathrm{f} } \right) ^2
  + \left( \frac{ \sigma _{G_{CL}} }{G_\mathrm{CL}} \right) ^2   
  + \left( \frac{\sigma _T}{T_\mathrm{AH}} \right) ^2
  +  \left( \frac{\sigma_k}{k} \right) ^2 
  +  \left( \frac{\sigma_{r_\mathrm{s}}}{r_\mathrm{s}} \right) ^2 .           \label{eq:PD_res_err}
\end{equation}
Note that the uncertainty of $ k $ and $ r_\mathrm{s} $ are assumed to be a constant value over the measurements of the calibration 
($ V_\mathrm{f} $, $ G_\mathrm{CL} $ and $ T_\mathrm{AH} $). 
Since $ k $ and $ r_\mathrm{s} $ could change from measurement to measurement, 
they should be determined at anytime the setup is calibrated. 
The largest value should be used as a modest estimation. 

The drift in the offset from the servo circuit changes the gradient of linear response due to the term of $ k $ in Eq.\,(\ref{eq:H_PD}). 
At the ideal operation point (mid-fringe), $ k = 1 $. 
The offset phase shift corresponding to the 0.3\,\% change in the slope is calculated by $ p = \arccos{(0.997)} = 0.078 $. 
This gives a normalized output voltage of 0.078 in the first order linear approximation.
This means that the 0.3\,\% criterion is satisfied if a drift of the offset can be kept around the ideal operation point within 7.8\,\% of $ V_\mathrm{pp} /2 $. 
The uncertainty is calculated by $ \sigma_k = 1 - \cos{(p)} $. 
When the offset is larger than 7.8\,\% of $ V_\mathrm{pp} /2 $, one should use the actual offset-phase, which can be derived from Eq.\,(\ref{eq:PD_MI_sin}).
The phase shift $ p $ is obtained from a output voltage of $ dV_\mathrm{PD} $ with a relation of $ \sin(p) = 2V_\mathrm{p}/V_\mathrm{pp} $ ($ V_\mathrm{p} \geq 0 $), 
where $ V_\mathrm{p} $ is the offset voltage. 
The deviation of $ k $ is expressed as $ k - \sigma_k = \cos{(p + E_\mathrm{p})} $ and $ k + \sigma_k = \cos{(p - E_\mathrm{p})} $, when $ 0 < p < \pi/2 $. 
Here, $ E_\mathrm{p} $ is upper and lower limit of the uncertainty of the measurement of $ p $. 
In order to satisfy the 0.3\,\% criterion, the measurement of $ p $ is according to the following two inequalities: 
\begin{equation}
\begin{array}{c}
 ( k - \sigma_k )/k = \cos{(p + E_\mathrm{p} )}/\cos{(p)} \geq  0.997       \\
 ( k + \sigma_k )/k = \cos{(p - E_\mathrm{p} )}/\cos{(p)} \leq  1.003      ,           \label{eq:Ep}
\end{array}
\end{equation}
when $ 0.078 < p < \pi/2 $. 
The equal signs of the above inequalities are illustrated in Fig.\,\ref{fig:TiltCorr}(b). 
This graph indicates the required accuracy for the measurement of $p$ to satisfy the 0.3\,\% criterion 
(for example, $p$ should be measured within an uncertainty of 0.01\,rad, when $p = 0.3$). 
The reason for the two curves in Fig.\,\ref{fig:TiltCorr}(b) can be understood as follows. 
In the case of a small offset, a negative deviation goes to the ideal operation region (the slope changes a little), 
and a positive deviation goes to a deviate region (the slope changes a lot). 
To stay within the 0.3\,\% criterion different precisions $ E_\mathrm{p} $ are required that can be derived from Eq.\,(\ref{eq:Ep}). 
In the case of a large offset, the requirement for the precision becomes more severe. 
By applying Type B evaluation, such requirement is mitigated by $ \sqrt{3} $. 
Actually, our measurements of the calibration factors, 
$ T_\mathrm{AH} $, $ G_\mathrm{CL} $ and $ V_\mathrm{f} $ have about 3\,\% of drift around the zero offset: 
$ ( \sigma_k / k ) ^2 = [ (1 - \cos{0.03}) / (\sqrt{3} \cos{0.03}) ] ^2 = ( 0.03\,\% ) ^2 $. 

Let us consider a nonlinear effect by a residual error signal. 
When the transfer functions are measured, a large signal is injected to the control loop to increase the signal to noise ratio.
In this situation a residual error signal may exceed the linear range of the mid-fringe, turning to be sinusoidal shape. 
In order to satisfy the criterion of 0.3\,\%, 
a condition of $ r_\mathrm{s} = \sin{(q)} / q \geq  0.997 $ ($ 0 \leq q \leq \pi/2 $) is imposed, which implies $ q \leq  0.13 $. 
Here, $ q $ is a phase shift from the operation point. 
The deviation of $ r_\mathrm{s} $ becomes $ \sigma _{r_\mathrm{s}} = 1 - \sin{(q)} / q $. 
For general offset $ p $, a phase shift $ q $ is obtained from the relation $ q \cos{(p)} = 2V_\mathrm{q}/V_\mathrm{pp} $, 
where $ V_\mathrm{q} $ is the residual error signal,  
and then, the inequality 
\begin{equation}
r_\mathrm{s} = \frac{\sin{(p + q)} - \sin{(p - q)}}{2 q \cos{(p)}} \geq 0.997
\end{equation}
is used as the 0.3\,\% criterion. 
This also sets a limit on $ q $ at 0.13 regardless of the parameter $ p $.
This means that the residual error signal $ V_\mathrm{q} $ should be suppressed within $ k \times (13\,\%) $ of the sinusoidal peak $ V_\mathrm{pp} /2 $. 
In our measurement, the peak to peak voltage is $ V_\mathrm{pp} = 3.44 $\,V and the standard deviation of the residual voltage is 0.094\,V. 
This is the largest value in all measurements of the calibration factors. 
It corresponds to 5.5\,\% of the sinusoidal peak and $ q = 0.055 $ by the small angle linear approximation. 
Using Type A evaluation, the contribution factor becomes 
$ ( \sigma _{r_\mathrm{s}} / r_\mathrm{s} ) ^2 = [ (0.055 - \sin{0.055}) / \sin{0.055} ] ^2 = ( 0.05\,\% ) ^2 $. 

\subsection{Calibration factors  $ V_\mathrm{f}, G_\mathrm{CL}, T_\mathrm{AH}, V_\mathrm{pp} $}
\begin{figure}[htb]
 \begin{center}
      \includegraphics[width=\linewidth ]{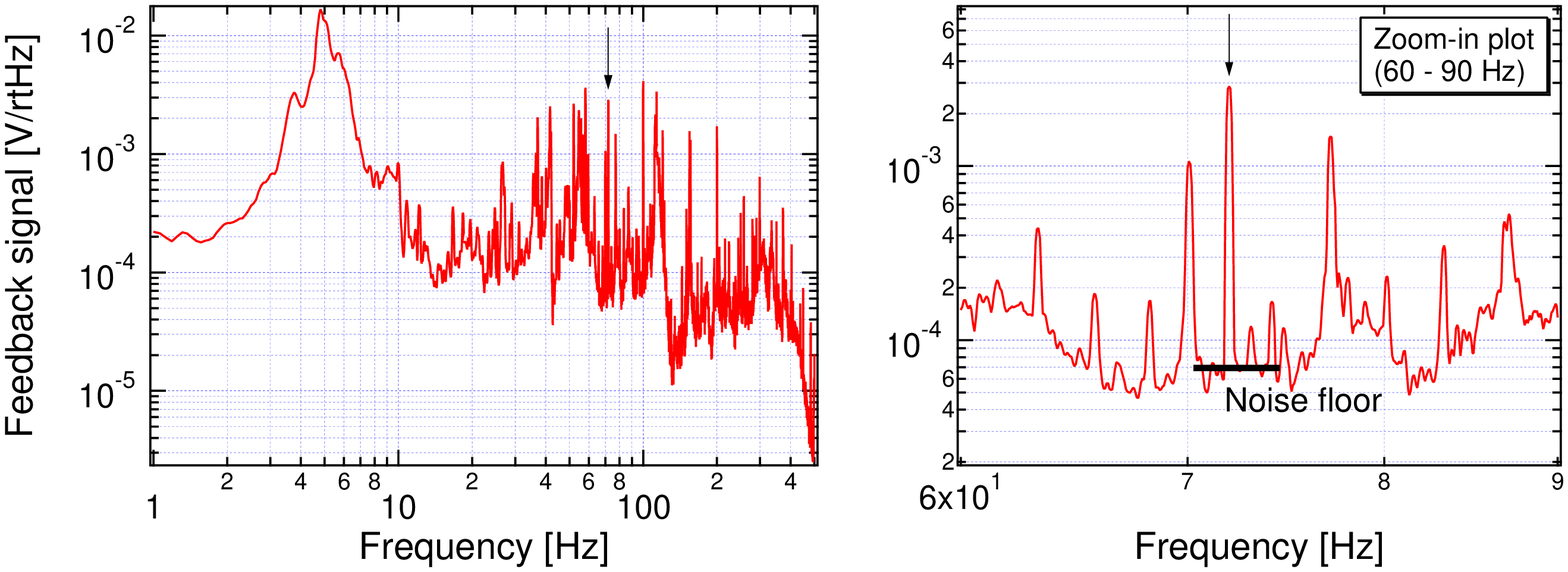}
         \caption{Feedback signal with modulated laser at 72\,Hz. Left: covering a wide frequency. 
Right: zoomed-in around 70\,Hz. The black arrow marks the peak at 72\,Hz. The black line indicates the noise level without the laser modulation. 
}
         \label{fig:FB}
  \end{center}
\end{figure}

Instead of the error signal from a PD, the feedback signal is used for the calibration factor of the displacement, 
so that the noise from the servo circuit can be suppressed. 
The data was averaged by using 31 FFTs based on data segments of 16\,s length. 
Figure\,\ref{fig:FB} shows the feedback signal in the case of the power modulation at 72\,Hz. 
There is an apparent peak at 72\,Hz, which is 40 times larger than the noise floor. 
The noise floor includes the readout noise and the residual motion of both the tiny mirror and large mirror due to disturbances 
like the seismic noise, acoustic vibration, detector noise, circuit noise and the frequency noise of the laser. 
These noises contribute to the peak value by $ (1/40)^2 \approx  0.06\,\% $. 
The measured standard deviation of the peak signal is 0.30\,\% of the peak value. 
This fluctuation includes the effects from the laser power like the intensity noise and thermal drift of the laser power. 
In total, the contribution term becomes $ ( \sigma _{V_\mathrm{f}} / V_\mathrm{f} ) ^2  = ( 0.30\,\% )^2 + ( 0.06\,\% )^2 $. 
The feedback signal was recorded at the same time as the output signal of the $ \mathrm{PD_{AOM}} $ 
to connect power fluctuation with the signal of $ \mathrm{PD_{AOM}} $ directly. 

\begin{figure}[htb]
 \begin{center}
      \includegraphics[width=\linewidth ]{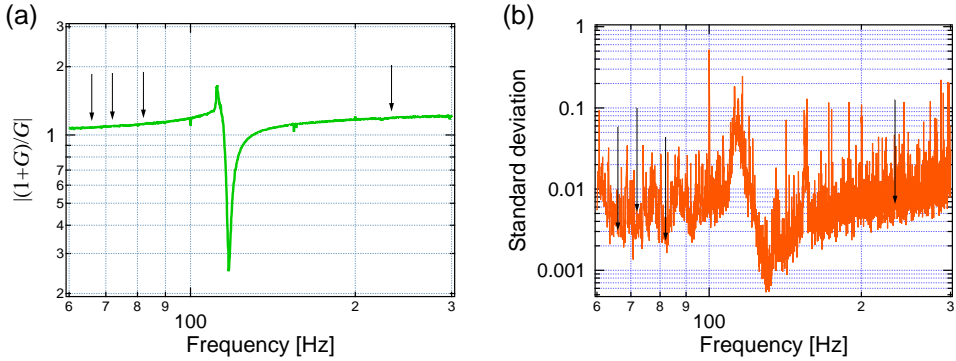}
         \caption{Analysis of the closed loop transfer function. 
(a) absolute value of the closed loop gain; 
(b) standard deviation of the closed loop gain. 
Black arrows indicate the measurement points. 
}
         \label{fig:CLG_StD}
  \end{center}
\end{figure}

Figure\,\ref{fig:CLG_StD}(a) shows a measurement of the closed loop gain $ G_\mathrm{CL} $.
A random noise was injected to measure this transfer function with a wide band of frequencies. 
As a result, the standard deviation is 0.48\,\% at 72\,Hz as shown in Fig.\,\ref{fig:CLG_StD}(b). 
The contribution term becomes $ ( \sigma _G / G_\mathrm{CL} ) ^2 = ( 0.44\,\% )^2 $. 
Also, the loop gain $ G $ was measured. 
The unity gain frequency is about 500\,Hz with a phase margin of 50 degrees. 

Since the transfer function $ T_\mathrm{AH} $ (from $ V_\mathrm{f} $ to $ dV_\mathrm{PD} $ in Fig.\,\ref{fig:Blockdiagram}) 
includes the response of the PD $ V_\mathrm{pp} $, 
both values should be measured without substantial delay to avoid any change of the conditions. 
If the uncertainty of $ V_\mathrm{pp} $ is increased, the uncertainty of $ T_\mathrm{AH} $ is also increased. 
This means that there exists a positive correlation between $ T_\mathrm{AH} $ and $ V_\mathrm{pp} $. 
By applying Eq.\,(\ref{eq:Sigma}) to Eq.\,(\ref{eq:Power_2}), 
the correlation term between $ T_\mathrm{AH} $ and $ V_\mathrm{pp} $ could yield a negative sign, which reduces total uncertainty, 
by the second term of Eq.\,(\ref{eq:Sigma})
because $ r( T_\mathrm{AH}, V_\mathrm{pp} ) $ is positive and the partial derivative of Eq.\,(\ref{eq:Power_2}) with regard to $ V_\mathrm{pp} $ is negative. 
To obtain the correlation coefficient, the simultaneous measurement of both parameters are needed. 
However, they cannot be measured at the same time because $ T_\mathrm{AH} $ is measured in lock whereas $ V_\mathrm{pp} $ is measured out of lock. 
\begin{figure}[ht]
 \begin{center}
      \includegraphics[width=\linewidth ]{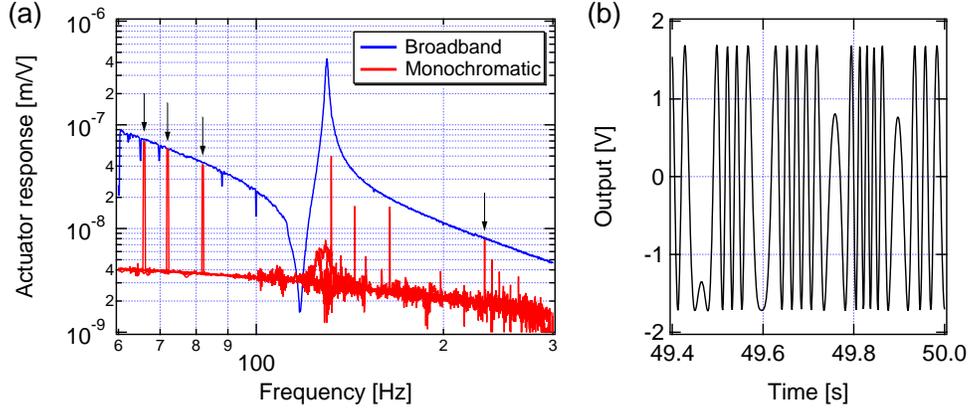}
         \caption{Analysis of the actuator response. 
(a) comparison of $ A $ between two measurement methods (monolithic and broadband), which are absolute values and divided by each PD response. 
Four monochromatic measurements are superimposed as the red solid line. Black allows are the measurement points; 
(b) output signal from the MI without lock. 
}
         \label{fig:ACT_Vpp}
  \end{center}
\end{figure}
This is why, they were treated as independent measurements to assume the worst case and Eq.\,(\ref{eq:Sigma_P}) is kept. 
Figure\,\ref{fig:ACT_Vpp}(a) shows the actuator response $ A = \lambda T_\mathrm{AH} / ( 2 \pi V_\mathrm{pp} ) $. 
The efficiency can be changed in various ways, for example, by changing the gap between coils and magnets, and by altering the balance of the four coils. 
These appear as the uncertainty of the measurement when the large mirror is shaken for the measurement of the transfer function. 
In order to reduce the uncertainty, the transfer function is measured at each monochromatic frequency that we select for uncertainty evaluation. 
A broadband measurement was also performed using random noise as the input signal, which has a somewhat large uncertainty (1.3\,\% - 1.9\,\%). 
It was consistent with the monochromatic measurements within their uncertainty. 
The resonance structure around 120\,Hz comes from a coupling with the angular motion of the large mirror. 
The broadband response is used for only making the displacement curve (Fig.\,\ref{fig:Sens}) but not used for the precise uncertainty evaluation.
The monochromatic measurement indicates $ ( \sigma _T / T_\mathrm{AH} )^2  = ( 0.49\,\% )^2 $ at 72\,Hz. 
This is the largest uncertainty of all measurement. 
Figure\,\ref{fig:ACT_Vpp}(b) is a part of measured output signal from the MI without lock. 
A fluctuation of the contrast of the MI appears in the change of the peak to peak value of the output $ V_\mathrm{pp} $. 
The peak to peak value is measured for 50\,s before and after measuring the transfer function $ T_\mathrm{AH} $. 
The values for maximum and minimum voltage of a fringe have been measured using sufficiently long segments of about 0.5\,s as covered in Fig.\,\ref{fig:ACT_Vpp}(b),
then the average and standard deviation are obtained using those data. 
The result is $ 3.437 \pm 0.011 $\,V and, subsequently, $ ( \sigma _V / V_\mathrm{pp} ) ^2   = ( 0.33\,\% )^2 $. 

There is a correlation between $ G_\mathrm{CL} $ and $ T_\mathrm{AH} $ because $ G_\mathrm{CL} $ includes $ T_\mathrm{AH} $ according to Fig.\,\ref{fig:Blockdiagram}; 
$ G_\mathrm{CL} = (1+G)/G = (1 + S T_\mathrm{AH})/ (S T_\mathrm{AH}) $. 
Since $(\partial{G_\mathrm{CL}} / \partial{G}) / G_\mathrm{CL} = -1/{[G(1+G)]}$, the relative uncertainty of $G$ is attenuated by $(1+G)$ in $ G_\mathrm{CL} $. 
If the origin of the relative uncertainty of $G$ comes from only $ T_\mathrm{AH} $, $ \sigma_G / G = \sigma_T / T_\mathrm{AH} = 0.49\,\% $ is attenuated by $(1+G)$. 
It means the correlation function is $r(G_\mathrm{CL}, T_\mathrm{AH}) = 1/(1+G)$. 
In this measurement, 
the relative correlation contribution becomes $ \{ 2 \times 0.44 \times 0.49 / (1+10) \}^2 = (0.04\,\%)^2 $ from Eq.\,(\ref{eq:Sigma}) due to $G = 10$ at 72\,Hz. 
According to this estimation, 
it is apparent that the relative contribution term of $ G_\mathrm{CL} $ (0.44\,\% at 72\,Hz) mainly comes from the uncertainty of the servo gain $S$.
The fluctuation of $S$ does not affect the measurement of $ T_\mathrm{AH} $ as far as its coherence is kept. 
Thus the maximum correlation is $1/(1+G)$.

\subsection{Mechanical response $H_{\mathrm{m}}$ }
\label{Sec:Hm}
In order to realize a quasi free-mass response of the 25\,mg mirror, it is suspended as a double pendulum with eddy current damping at the intermediate mass. 
Actually, the suspended mass is regarded as almost a free mass in the high frequency region ($ H_{\mathrm{m}} \approx  1/(m \omega ^2) $).
Although parameter deviations of the the suspension system have the strongest impact on $ H_{\mathrm{m}} $ around the suspension resonance, 
they also influence the mechanical transfer function even at higher frequencies. 
Based on a mechanical model these deviations are investigated in the following. 
Figure\,\ref{fig:DoublePend}(b) shows a model of the double pendulum. 
The equation of motion is written as 
 \begin{equation}
                   m_1 \ddot{x} _1 = - \frac{ (m_1 + m_2) g }{ l_1 }  x_1 - \frac{ m_2  g }{ l_2 } (x_1 - x_2) - \Gamma _1 \dot{x} _1  ,
 \end{equation}
 \begin{equation}
                   m_2 \ddot{x} _2 = - \frac{ m_2 \, g }{ l_2 } (x_2 - x_1) - \Gamma _2 \dot{x} _2 + F ,
 \end{equation}
where 
$ m $ is the mass, 
$ x $ the displacement, 
$ l $ the length of the suspension wire, 
$ \Gamma $ the damping coefficient of the eddy current or air, 
$ g $ the acceleration due to gravity at the Earth's surface and 
$ F $ the force from the radiation pressure. 
The subscripts of 1 and 2 denote the upper pendulum and lower pendulum, respectively. 
After the Fourier transformation, these equations are rewritten as 
 \begin{equation}
 A
   \left(
  \begin{array}{c}
        \tilde{x} _1 \\
        \tilde{x} _2
  \end{array}
   \right)
=
   \left(
  \begin{array}{c}
      0     \\
 \tilde{F}
  \end{array}
   \right) , \hspace{1cm}
 A ^{-1}
\equiv 
   \left(
  \begin{array}{cc}
      a_{\mathrm{m}} & b_{\mathrm{m}}     \\
      c_{\mathrm{m}} & d_{\mathrm{m}}     
  \end{array}
   \right).
\end{equation}  \label{eq:Matrix} 
Here, the matrix $A$ includes complex forms of above parameters.
The response of the test mass becomes 
\begin{equation}
 H_{\mathrm{m}} (\omega ) \equiv \frac{ \tilde{x} _2 }{ \tilde{F} } = d_{\mathrm{m}}  .
\end{equation}

\begin{figure}[tb]
 \begin{center}
      \includegraphics[width=\linewidth ]{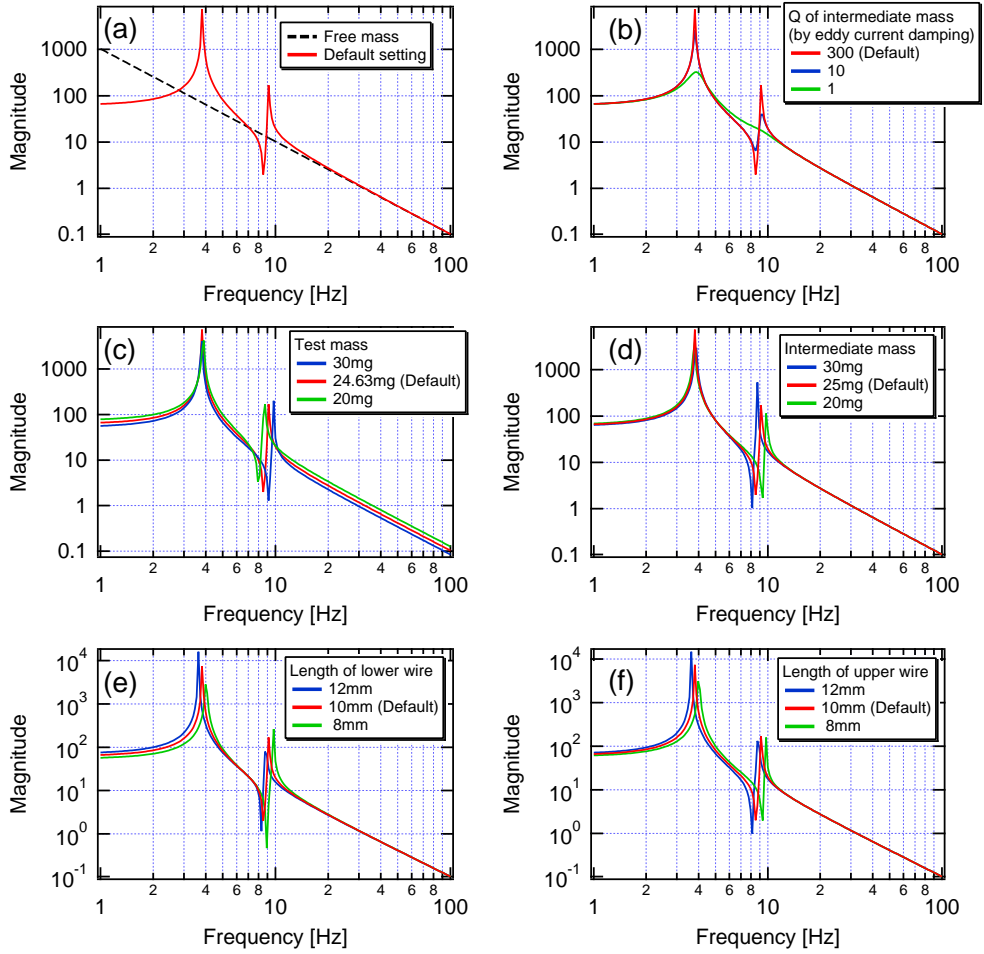}
         \caption{Effects from parameter deviations of the double pendulum on its mechanical response. 
         (a) comparison between the free mass response and mechanical response with the default setting; 
         (b-f) deviations from the default setting in case the possible differences exist in the parameters of  
                the damping factor ($ Q $) at the intermediate mass,  
                the test mass, 
                the intermediate mass, 
                the length of the upper wire, 
                and of the lower wire, respectively.
}
         \label{fig:MechRes}
  \end{center}
\end{figure}
The default-setting parameters of our small pendulum are $ m_1 = 25 $\,mg, $ m_2 = 24.63 $\,mg, $ l_1 = 10 $\,mm, $ l_2 = 10 $\,mm and $ Q = 300 $. 
The quality factor $ Q $ is related with the damping coefficient by $ \Gamma _1 = m_1 \omega _1 / Q $ using $ \omega _1 = \sqrt{ g / l_1 } $.
The quality factor of 300 is the measured value of another single pendulum with 25\,mg mirror and 10\,$\mu $m suspension wire in air.
This is the highest value in the case as the middle mass is damped by only air.
The quality factor of the lower pendulum is fixed at 300 with regard to $ \Gamma _2 $.
Figure\,\ref{fig:MechRes}(a) shows the default setting of the mechanical response $ H_{\mathrm{m}} $, 
comparing with that of the free mass $ 1/(m_2 \, \omega ^2) $. 
The differences found are 0.58\,\%, 0.48\,\%, 0.37\,\% and 0.05\,\%, at 66, 72, 82 and 233\,Hz, respectively. 
These are small but quite large compared with the requirement. 
To avoid such deviation effect due to the suspension system, 
we use the default-setting response (red solid line of Fig.\,\ref{fig:MechRes}) as the mechanical response $ H_{\mathrm{m}} $ instead of the free mass response. 
Also, the quasi free-mass response is ensured by the $f ^{-2}$ response of the peaks in Fig.\,\ref{fig:Sens}.
It means that there is no strange and large deviation from the expected response at the observation frequencies. 

The response $ H_{\mathrm{m}} $ has still some components of uncertainty due to an imperfection in making the double pendulum.
Figure\,\ref{fig:MechRes}(b-f) show the impact of deviations from the mechanical parameter on the mechanical response $ H_{\mathrm{m}} $. 
The estimation ranges are set for the wire lengths $l_1$ and $l_2$ to be $ \pm 2 $\,mm, 
for the intermediate mass $m_1$ to be $ \pm 5 $\,mg and 
for the $ Q $ from 1 to 300. 
Type B evaluation is used for these upper and lower limits. 
As expected, marked differences can be seen at a region near the resonance. 
The contributions from each uncertainty at 72\,Hz are 0.001\,\% from $m_1$, 0.00001\,\% from $l_1$, 0.12\,\% from $l_2$ and 0.00001\,\% from $Q$.
The imperfection effect from the length of the lower fiber has relatively large effect. 
It is apparent that an uncertainty of the test mass, 
$ m_2 $, causes a large effect over whole frequency region directly as shown in Fig.\,\ref{fig:MechRes}(c).
The weight of the 25\,mg mirror have been estimated by measuring its dimensions (the diameter and thickness) using slide calipers.
The actual value is $ 24.63 \pm 0.16 $\,mg, corresponding to 0.65\,\% uncertainty. 
From the above discussion, the total contribution of the uncertainty becomes  
$(\sigma _H / H_{\mathrm{m}} ) ^2 \approx (0.12/\sqrt{3})^2 + (0.65/\sqrt{3})^2 = (0.38\,\%)^2 $. 

\subsection{Transfer efficiency of momentum $ \alpha _\mathrm{r} $ } 
The reflection, scattering and absorption have different transfer efficiencies of the momentum from photons to the mirror. 
The transfer efficiency of the reflection is $ 2 \Delta P $, as $ \Delta P$ is the momentum of one photon. 
The scattering has a range from $ \Delta P$ to $ 2 \Delta P $ because of the uncertainty in the scattered direction. 
On the other hand, the absorption is $ \Delta P$. 
The total transfer efficiency of momentum $ \alpha _\mathrm{r} $ is measured including these uncertainty. 
The incident, reflected and transmitted laser powers were measured by repeating ten times 
using a commercial power meter (Coherent PowerMax, PS19Q) that has 2.5\,\% uncertainty according to the specification sheet. 
Thus the uncertainty of the reflectivity becomes $ 2.5\,\% \times (\sqrt{2} / \sqrt{10}) = 1.12\,\% $ 
where the $ \sqrt{2} $ comes from the two power measurements of the incident and reflected. 
The reflectivity, transmittance and loss are 
$ R = 0.9941 ^{+0.0053} _{-0.0111} $, 
$ T = (657 \pm 5) \times 10^{-6} $, 
and $ L = 0.0053^{+0.0111} _{-0.0053}$, respectively. 
Here the relation of $ R + T + L = 1 $ and the law of the conservation of energy are assumed. 
Since we cannot distinguish the origin of the loss from the scattering and absorption, the upper and lower limits are estimated. 
The upper limit of the transfer efficiency $ \alpha _\mathrm{r} $ becomes 0.9993 by assuming the no loss material. 
The lower limit of it is derived from $ R + L/2 = (0.9941 - 0.0111) + (0.0053 + 0.0111)/2 = 0.9912 $ by assuming the origin of the loss is absorption only. 
Hence the transfer efficiency is $ \alpha _\mathrm{r} = 0.9952 \pm 0.0041 $ as the center of this range. 
This results in $ ( \sigma _{\alpha } / \alpha _\mathrm{r} ) ^2  = ( 0.24\,\% )^2  $ using Type B evaluation. 

The beam spot of the modulated laser has a diameter of 0.4\,mm on the tiny mirror. 
The reflection area is sufficient because the reflection coating covers 80\,\% of the surface (2.4\,mm) on the tiny mirror.
Even if there is an off center of the beam spot, the reflection area can be kept to 1\,mm at least. 
In this case, the loss of the power becomes 0.0004\,\% and is negligible. 

\subsection{Measurement of incident angle  $\phi $ }
The incident angle $\phi $ is projected on the horizontal direction $ \phi_\mathrm{H} $ and vertical direction $ \phi_\mathrm{V} $ to measure the angles, 
i.e. $ \cos{\phi } \equiv  \cos{\phi_\mathrm{H} }\cos{\phi_\mathrm{V} } $.
The lengths of $L1$ and $L2$ in Fig.\,\ref{fig:Setup} were measured using a ruler to decide the incident angle of the horizontal direction. 
The results are $L1 = 388 \pm 1$\,mm and $L2 = 202 \pm 1$\,mm, whose uncertainty stem from the minimum unit of the ruler. 
It corresponds to the angle $ \phi_\mathrm{H} =  0.481 \pm 0.004 $\,rad. 
This kind of measurement is classified as Type B evaluation. 
The contribution term for the standard uncertainty is 
$ \sigma _{\phi_\mathrm{H}} \tan \phi_\mathrm{H} =  \left( 0.004/\sqrt{3} \right) \tan{0.481} = 0.11\,\% $. 
In addition to the horizontal direction, the vertical angle was also measured to be $ \phi_\mathrm{V} = 0.046 \pm 0.002 $\,rad. 
The contribution term is $ \sigma _{\phi_\mathrm{V}} \tan \phi_\mathrm{V} = 0.004\,\% $. 
The horizontal term is dominant because of its large angle of $ \phi_\mathrm{H} $. 

\subsection{Rotational effect  $ R_\mathrm{c} $ }
\label{Sec:Rc}
The actual mirror is not a point mass but a rigid body. 
This gives additional rotational degrees of freedom.
If a laser beam hits the mirror surface off center, the mirror rotate around its center of mass because of the induced torque. 
This rotational motion is added to the translational direction by a product of the rotational angle and the off-centering distance. 
The observation spot on the mirror surface is also important if a laser interferometer is used as a displacement sensor. 
This coupling effect from the mirror rotation to the longitudinal direction of the MI was named rotational effect above. 
The induced angular tilt by the incident laser power can change the contrast of the MI. 
However, the position of equilibrium was not changed further since the DC laser power was kept constant during all measurements (with and without modulation). 
Thus, the rotational effect arises solely from the power modulation.

\begin{figure}[hu]
 \centering
      \includegraphics[width=5.5cm ]{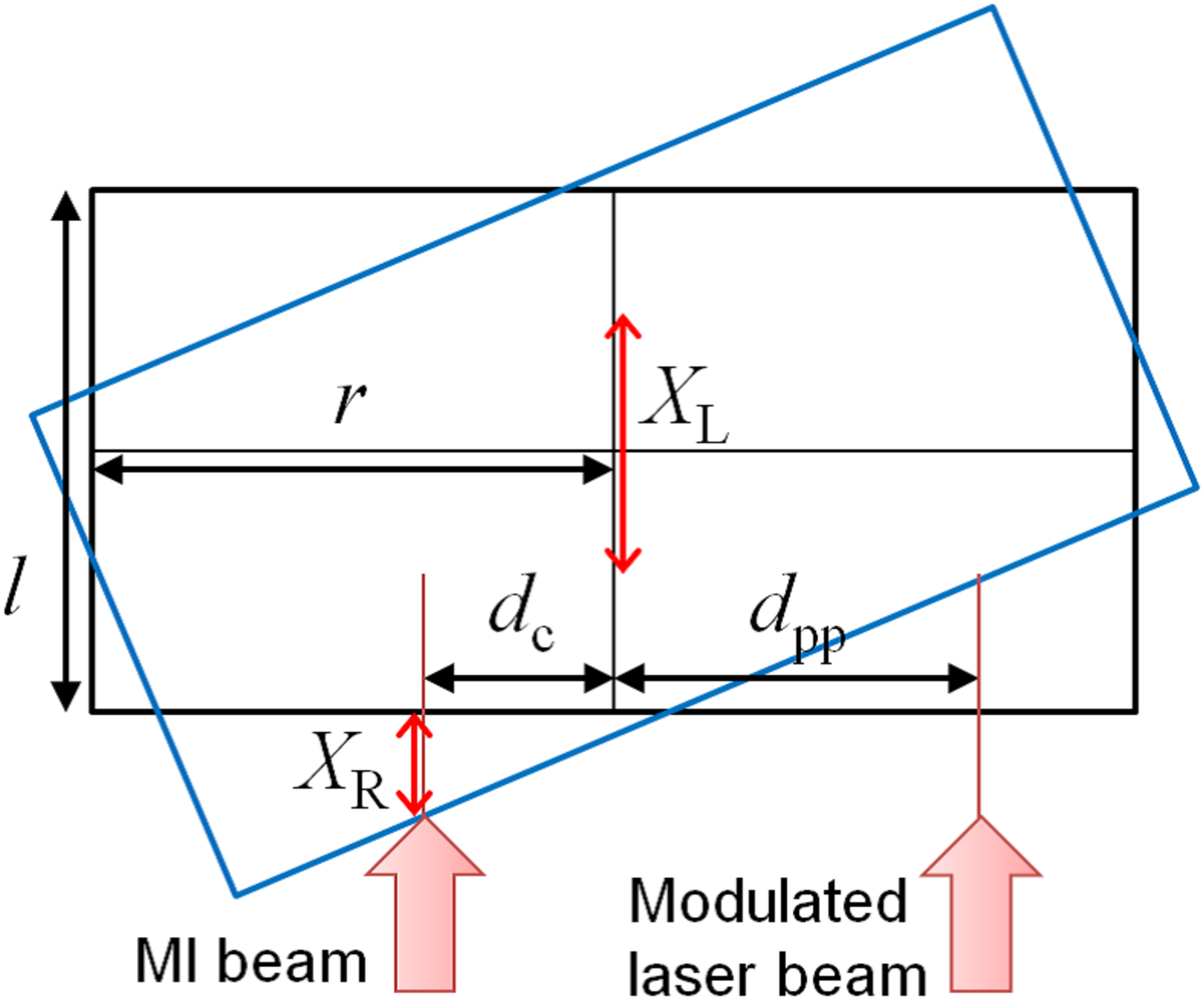}
         \caption{Illustration of the rotational effect. The red arrows show the displacement of the mirror. 
}
         \label{fig:Rotation}
\end{figure}

The rotational effect is defined as a displacement ratio between the translation and rotation caused by the radiation pressure \cite{Hild_phD}. 
It is written as $ R_\mathrm{c} = X_\mathrm{R} / X_\mathrm{L} = \pm (m d_\mathrm{c} d_\mathrm{pp})/I  $ as can be derived based on Fig.\,\ref{fig:Rotation}. 
Here, 
$ X_\mathrm{L} $ is the displacement for the longitudinal direction with respect to MI due to the radiation pressure, 
$ X_\mathrm{R} $ is the longitudinal displacement caused by the rotation, 
$ m $ the mass of the mirror, 
$ I $ the moment of inertia given by $ I = m (r^2/4 + l^2/12) $, 
$ r $ the mirror radius, 
$ l $ the thickness of the mirror, 
$ d_\mathrm{c} $ the off-centering distance of the beam spot of the MI on the tiny mirror and 
$ d_\mathrm{pp} $ the distance of off-centering spot of the intensity modulated laser. 
The sign shows the relation of the beam spot positions between the MI and intensity modulation. 
In order to suppress the rotational effect, precise adjustment of the beam centering is needed.
If both off-centering distances are comparable ($ d_\mathrm{c} = d_\mathrm{pp} = d_\mathrm{0} $), 
the inequality becomes simply $ (m d_\mathrm{0} ^2)/I \leq 0.003 $ for the 0.3\,\% criterion, which has solution $ d_\mathrm{0} \leq 47$\,$\mu $m.
The rotational effect can be suppressed below 0.3\,\% uncertainty when the beam centering is adjusted within about 50\,$\mu $m including its uncertainty. 
By implementing Type B evaluation, the requirement is extended up to 80\,$\mu $m. 
In section\,\ref{Sec:QE}, there was a discrepancy between the measurement result and the value from the specification sheet of the photodiode. 
This can be explained by the rotational effect with a beam off-centering of about $d_\mathrm{0} = 400$\,$\mu $m 
when other effects (temperature and uniformity of PD surface) are not taken into account.
If the off-centering distances are quite large, a finite value of $ R_\mathrm{c} $ is used and its uncertainty should be evaluated. 
The uncertainty of the rotational effect is expressed as 
\begin{equation}
\sigma _R = 
\sqrt{
2 \left[ \frac{d_\mathrm{0}}{(r^2/4 + l^2/12)} \sigma _d \right] ^2 + 
\left[ \frac{d_\mathrm{0} ^2}{(r^2/4 + l^2/12)^2} \frac{r}{2} \sigma _r \right] ^2 + 
\left[ \frac{d_\mathrm{0} ^2}{(r^2/4 + l^2/12)^2} \frac{l}{6} \sigma _l \right] ^2
} ,
\label{eq:Rc_err}
\end{equation}
where $ \sigma _d $, $ \sigma _r $ and $ \sigma _l $ are the measurement uncertainties of $ d_\mathrm{0} $, $r$ and $l$, respectively.
If the slide calipers are used for measuring the dimension of the tiny mirror, 
the second and third term become negligible compared to the first term. 
Also, the beam position of the intensity modulation can be adjusted precisely by comparing it with the MI. 
The first term should be unfolded for each beam position. 
Then the contribution term becomes 
\begin{equation}
\left[ \frac{\sigma _R}{ (1 + R_\mathrm{c}) } \right] ^2 
\approx  
\frac{1}{ (1 + R_\mathrm{c})^2 }
\left\{
  \left[ \frac{d_\mathrm{c}}{(r^2/4 + l^2/12)} \frac{ \sigma _{d_\mathrm{c}} }{\sqrt{3} } \right] ^2 + 
  \left[ \frac{d_\mathrm{pp}}{(r^2/4 + l^2/12)} \frac{ \sigma _{d_\mathrm{pp}} }{\sqrt{3} } \right] ^2 
\right\}  ,
\end{equation}
where the factor of $ \sqrt{3} $ is the coefficient from Type B evaluation. 
The position uncertainty of 10\,$ \mu $m is obtained from one pixel when the area of $1 \times 1$\,$\mathrm{cm}^2$ is photographed using a CCD camera with 1\,M pixel.
If we adjust the beam centering of the intensity modulation within $d_\mathrm{pp} = (100 \pm 10)$\,$ \mu $m, the uncertainty of this term becomes 0.08\,\%. 
The dominant part is the beam position of the MI because it is more difficult to adjust beam centering with keeping a good contrast. 
In order to achieve the 0.3\,\% criterion, 
the centering of the beam from the interferometer $ d_\mathrm{c} $ needs to be adjusted within $ (390 \pm 10)$\,$ \mu $m. 
This is achievable. 
Incidentally, in the case of a modest requirement (measurement uncertainty of 20\,$ \mu $m), 
the contribution part of $ d_\mathrm{pp} $ becomes 0.15\,\% by assuming the beam centering of the power modulation with $(100 \pm 20)$\,$ \mu $m 
and the uncertainty of $ d_\mathrm{c} $ is kept on 0.30\,\% by assuming the centering of the MI with $(200 \pm 20)$\,$ \mu $m. 
These are also feasible values. 

A rotation of the tiny mirror can change the reflected-beam angle for the MI. 
This effect is included in the uncertainty evaluation of $ V_\mathrm{f} $ as a fluctuation of the peak signal. 
Also, if the two beam spots on the tiny mirror are adjusted to the same position, 
a local deformation of the mirror could appear by the radiation pressure \cite{Hild2007}. 
It is omitted because that effect appears in a high frequency region compared with our measurement region. 

\section{Conclusion and discussion} 
We have proposed an optomechanical power meter, which can count the number of photons by measuring the displacement of a mirror pushed by a modulated laser beam. 
As a prototype test, a suspended 25\,mg mirror is used for an optomechanical-coupled oscillator and its displacement is measured using the Michelson interferometer. 
Table\,\ref{tbl:Budget} shows the summary of the uncertainty contribution. 
The total standard uncertainty is 0.92\,\% plus the quadrant sum of a term from the rotational effect, which is not measured but discussed. 
Hence, we do not claim the 1\,\% accuracy in this prototype experiment because of the lack of the beam centering information. 
If the uncertainty due to the rotational effect is assumed to be 0.30\,\% for $ d_\mathrm{c} $ and 0.15\,\% for $ d_\mathrm{pp} $, the total uncertainty becomes 0.98\,\%. 
When $ P_\mathrm{m} $ is converted to the DC power, the uncertainty of $ \mathrm{PD_{AOM}} $ is added. 
The measured contribution term is $ (\sigma_{V_\mathrm{m}} / V_\mathrm{m})^2 = (0.006\,\%)^2 $ at 72\,Hz, which is negligible. 
According to our demonstration and calculation analysis, 
we can conclude that it is achievable for our new kind of power meter to measure the laser power within one percent uncertainty (1\,$\sigma $). 
Furthermore, average of four comparable measurements (at 66, 72, 82 and 233\,Hz in this case) improve the uncertainty by $ \sqrt{4} $, 
i.e. 0.5\,\% uncertainty is obtained. 
Each measurement can be regarded as independent measurement since main contributions of the uncertainty come from the frequency specific calibration factors. 
\begin{table}[b]
\begin{center}
\caption{Uncertainty budget of the prototype experiment at 72\,Hz.}
\label{tbl:Budget}
\vglue5pt
 \begin{tabular}{rlrr}                                                                    
\hline
\multicolumn{1}{r}{Notation} & \multicolumn{1}{l}{ } & \multicolumn{1}{l}{Type} & \multicolumn{1}{r}{Relative contribution} \\ 
\hline
   & \bf{Displacement sensor} & & \\
\hline                           
  $dV_{\mathrm{PD}}$     & Deviation from mid-fringe      &   &  \\
  $ k $                  & Offset and drift                        & B & $  0.03\,\%  $  \\ 
  $ r_\mathrm{s} $       & Residual error signal                   & A & $  0.05\,\%  $  \\
\hline
  $dV_{\mathrm{PD}}$     & Calibration factors            &   &  \\
  $ V_\mathrm{f} $       & Feedback signal (peak value)            & A & $  0.30\,\%  $  \\
                         & Feedback signal (noise floor)           & A & $  0.06\,\%  $  \\
  $ G_\mathrm{CL} $      & Closed loop gain                        & A & $  0.44\,\%  $  \\
  $ T_\mathrm{AH} $      & Transfer function from actuator to PD   & A & $  0.49\,\%  $  \\
  $ V_\mathrm{pp} $      & Peak to peak of MI                      & A & $  0.33\,\%  $  \\
  $ (G_\mathrm{CL}, T_\mathrm{AH}) $      & Correlation term       & A & $  0.04\,\%  $  \\
\hline
   & \bf{Optomechanical response} & & \\
\hline                           
  $H_{\mathrm{m}}$       & Mechanical response            &   &  \\
  $ l_2 $                & Lower-fiber length                      & B & $  0.07\,\%  $  \\
  $ m_2 $                & Mass of tiny mirror                     & B & $  0.38\,\%  $  \\
\hline
  $ \alpha _\mathrm{r} $ & Transfer efficiency of momentum         & B & $  0.24\,\%  $  \\
\hline
  $ \phi_\mathrm{H} $    & Incident angle (horizontal)             & B & $  0.11\,\%  $  \\   
\hline
  $ R_\mathrm{c} $       & Rotational effect              &   &  \\
  $ d_\mathrm{c} $       & Beam centering of MI                    & B & $ \frac{1}{ 1 + R_\mathrm{c} }
  \left[ \frac{d_\mathrm{c}}{(r^2/4 + l^2/12)} \frac{ \sigma _{d_\mathrm{c}} }{\sqrt{3} } \right]  $  \\
  $ d_\mathrm{pp} $      & Beam centering of intensity modulation  & B & $ \frac{1}{ 1 + R_\mathrm{c} }
  \left[ \frac{d_\mathrm{pp}}{(r^2/4 + l^2/12)} \frac{ \sigma _{d_\mathrm{pp}} }{\sqrt{3} } \right]   $  \\
\hline
\multicolumn{4}{l}{Total} \\
\multicolumn{4}{c}{
$ \frac{\sigma _P}{P} = 
\sqrt{
\frac{1}{ (1 + R_\mathrm{c} )^2 }
\left\{ 
\left[ \frac{d_\mathrm{c}}{(r^2/4 + l^2/12)} \frac{ \sigma _{d_\mathrm{c}} }{\sqrt{3} } \right ]^2 + 
\left[ \frac{d_\mathrm{pp}}{(r^2/4 + l^2/12)} \frac{ \sigma _{d_\mathrm{pp}} }{\sqrt{3} } \right ]^2
\right\} +
( 0.92\,\% )^2
} 
$ }  \\
\hline
 \end{tabular}
 \end{center}
\end{table}

There are some means to improve this measurement. 
Commercial precision weighing balances can measure the weight of mass with 0.01\,\% uncertainty \cite{Balance}.
The uncertainty of the closed loop gain is expected to be reduced by the monochromatic measurement. 
By moving the position of the AOM to the MI light way, the intensity noise is canceled using the subtraction of the symmetric port from the antisymmetric port. 
It means that the uncertainty of the feedback signal can be reduced and the incident angle vanishes. 
The uncertainty of the actuator response can be reduced by adjusting the gap between coils and magnets to the insensitive region of efficiency, 
and also making a good suspension of the actuation mirror. 
Increasing the number of reflectivity measurements reduces the uncertainty in the transfer efficiency of momentum. 
A short length MI, including the path up to a PD, is favorable to stabilize the contrast. 
Averaging the measurements at other frequencies can improve the precision, 
thus a refined setup can be expected to reach the comparable uncertainty of the current primary standard. 

The available power of this apparatus is limited by specific conditions. 
The maximum power is limited by the damage threshold of the coating of the oscillator (tiny mirror). 
If a higher power is needed to measure properly, the mirror mass should also be increased to keep the response of the displacement sensor linear. 
The minimum power is limited by the signal to noise ratio of the radiation pressure response. 
The available wavelength depends on the reflecting or absorbing condition of the oscillator. 
In this paper, our mirror coating is optimized to reflect a laser beam with a wavelength of 1064\,nm. 
This reflection coating is changeable and should be optimized for the laser wavelength that is to be measured. 
To measure the X-ray region, the absorbing material is better to use though the transfer efficiency of the momentum is reduced by a factor of two in such case. 
For the radio wave region, a metal material can be used for reflection. 
Of course, it is important to investigate the transfer efficiency of the momentum precisely. 
To determine the DC power of the incident laser, 
we need to determine the modulation index of the power modulator or to prepare the reference receiver like $ \mathrm{PD_{AOM}} $. 
In case a reference receiver is used, the available power and wavelength depend on the properties of the receiver. 
Optical choppers can be substituted for the AOM to generate the power modulation. 

We have displayed a new method using an optomechanical response to measure the laser power via a displacement of an oscillator. 
In our technique, 
each parameter can be measured/estimated precisely although there are many parameters contributing to the uncertainty as shown in table\,\ref{tbl:Budget}. 
Even with this prototype experiment, no vacuum or cryogenic instruments are needed to obtain the level of one-percent uncertainty. 
By adjusting the mirror material and coating, it can be applied to an arbitrary laser wavelength. 
These characteristics can expand the application range of power meters. 
This method could be used not only to determine a primary standard, but also to make a new kind of commercial power meter in the future. 

\section{Acknowledgments}
We are grateful to Keita Kawabe, Stefan Hild and Koji Arai for many comments and helpful discussions. 
Also, we thank Joris van Heijningen for a careful reading. 
S. Kawamura was supported by a Grant-in-Aid for Scientific Research from the Japan Society for the Promotion of Science, 
G. DeSalvo was supported by the U.S. National Science Foundation under the University of Florida REU program. 
S. Sakata was supported by Young Scientists B. 

\end{document}